\documentclass[review]{elsarticle}
\usepackage[version=3]{mhchem} 
\usepackage{amsmath}

\usepackage{hyperref}

\journal{Fuel}

\begin{document}

\begin{frontmatter}

\title{Heteroatomic jet fuel components: Lichen substances as fuel component and potential additives}

\author{Pooja Sharma}
\address{Aerospace Studies, University of Toronto, Toronto, Canada}


\begin{abstract}

This article presents chemical analysis of jet fuel (Jet A-1) heteroatomic fuel components with identification of an antioxidant lichen acid in methanol extracted fuel samples. Thermal stressing of jet fuel produces soluble macromolecular oxidatively reactive species (SMORS) and heteroatomic deposits. SMORS are deposit precursors and elementary heteroatomic units containing unsaturated and aromatic hydrocarbons. Fuel additives such as antioxidants can inhibit SMORS and deposit formation within limited heating residence time and temperature range. Jet A-1 was thermally stressed in the autoxidation regime (150 to 300 \ce{^0}C) followed by spectroscopic analysis. Thermally stressed jet fuel static tests mass spectra show higher molecular weight compounds and SMORS in the mass range 300-1000 Da compared with unstressed fuel samples supporting deposition. Jet A-1 samples were analyzed by electrospray ion source mass spectrometry (ESI-MS), Fourier transform infrared (FTIR) and \ce{^${13}$}C nuclear magnetic resonance (NMR) spectroscopy. FTIR bands for oxygen containing species reveal the presence of alcohol, phenol and ether groups. NMR \ce{$^{13}$C } standard and distorsionless enhancement polarization transfer (DEPT 135) spectra show heteroatomic alkoxy species in both unstressed and thermally stressed fuel samples. Natural products polyphenols and lichen derived oxygenated compounds are excellent antioxidants. A new perspective of using lichen substances as fuel additives emerged in this study. Exploring further, natural products extraction methods optimization remains a key challenge and advantages of polyphenolic lichen acids as potential fuel and chemical additives are discussed.

\end{abstract}


\end{frontmatter}


\section{Introduction}

Jet fuel is a complex organic mixture of several hydrocarbons with many additives added in trace amounts to improve properties such as thermal stability, lubricity, anti-static, anti-icing and corrosion inhibition. Jet fuel additives can be categorized as hydrocarbon diluents and active ingredients. Static dissipater additives and lubricity improvers are types of hydrocarbon diluents. Active ingredients include i) antioxidants, ii) metal deactivators (MDA), iii) icing inhibitors or fuel system icing inhibitors and leak detection additives. Fuel antioxidants contain phenols, thiols and arylamines with weak O-H, S-H and N-H bonds strengths. Heteroatomic species such as phenols, indoles, pyrroles (2,5 dimethylpyrrole) and carbazoles are deposit precursors. Although undesirable for thermal stability, heteroatomic species enhance the lubricity of jet fuel which supports trace levels of N,O, S containing species present in the fuel. However, refining processes such as hydrotreatment remove heteroatomic species such as sulfur containing compounds. Few mono and di substituted heteroatomic deposit precursors are quinones, hydroquinones, 2,5 dimethylpyrrole (2,5 DMP), thiophene, benzothiophene, dibenzofuran, indoles and quinoline. Dissolved oxygen (DO) supports hydroperoxide oxidation reactions and oxygen removal can improve thermal stability \cite{Hazlett1991}. Thermal degradation of fuel hence can depend on compositional, process and operational factors such as longer heating residence time, dissolved oxygen, additives, crude, refinery treatment and distribution or routing of fuel in compact zones.
\\*

Lichen secondary metabolites are useful natural products with antioxidant, antimicrobial, anti-inflammatory and anticancer properties. Lichens are plant like organisms similar to lower plants primarily composed of fungi combined with algae, cynobacteria and have similar nutritional and symbiotic pattern. Lichen metabolites or lichen acids are produced extracellularly and deposited as crystals over fungi hyphae varying from 0.1 to 10 total weight percentage and contain weak phenolic acids such as depsides and depsidones \cite{Rankovic2015,Hawksworth1976}.  Lichens grow in the extreme environments such as Antarctica and old growth forests of North America, and their survival in such complex climate is supported by the presence of various compounds such as antioxidants which reduce oxidative stress. In biological systems, reactive oxygen species (ROS) along with similar reactive nitrogen and sulphur species can harm other molecules such as proteins, RNA and DNA by creating oxidative stress. Similarly, thermal stressing of hydrocarbon fuel generates fuel free radicals (FFRs) which can include \ce{R^{.}}, \ce{H^{.}}, \ce{ROO^{.}} and \ce{OH^{.}} (where R indicates alkyl or hydrocarbon chain) which can initiate multiple reactions resulting in solids formation in the fuel. Deposition as a result of thermal and storage instabilities accelerated by FFRs can change jet fuel composition and properties which is undesirable for aircraft gas turbine engine operation. Suppression of oxidation reactions can be accomplished by antioxidants which scavenge free radicals and control oxidation.
\\*

Heteroatomic species in the middle distillate fuels are present mostly as mono or di substituted cyclic and aromatic compounds which contribute in the higher mass species formation through reactions favoured by heating and oxidation. Hydrocarbon species in this class of compounds include quinones, indoles, pyrroles, benzofuran, thiophene, PAHs and other similar species. Polar hydrocarbon jet fuel species are extractable in methanol (MeOH) and other organic solvents such as tetrahydrofuran and isopropanol (IPA). Also, as reported by Hardy and Wetcher\cite{Hardy1990} SMORS are methanol extractable and possibly contain multiple unsaturated and aromatic molecules in the mass range more than 350 Da. MeOH is also a suitable extraction solvent for polyphenolic natural product antioxidants and a similar lichen acid is observed in the Jet A-1 samples analyzed in this study. 
\\*

SMORS are observed in the thermally stressed Jet A-1 fuel samples in the flow and static tests mass spectra in this work. Analysis by FTIR and NMR also confirmed oxygen containing compounds as jet fuel components and possibly contributing in the SMORS formation. An antioxidant lichen substance, gyrophoric acid is identified in the jet fuel spectroscopic analysis in this research. Possibility of lichen substance as a fossil fuel component with its  natural origin can be due to their growth in the complex climates and long ages of survival. Lichens growth rate is very slow and species with ages of upto 4500 years are found in Britain and Greenland \cite{Hawksworth1976}. Lobaria linita and Lobaraia pulmonaria are lichens which produce gyrophoric acid and found in old growth forests of North America and British Columbia (Canada). Lichen substances are excellent antioxidant and antimicrobial agents and their potential uses as fuel additives based on jet fuel samples analyzed in this study are discussed. In summary, spectroscopic analysis of SMORS in the thermally stressed Jet A-1 samples compared with unstressed fuel samples and a lichen acid is identified as an antioxidant jet fuel component supporting a new perspective of potential natural product fuel additives is presented in this paper.


\section{Methods}

Single tube heat exchanger is used for Jet A-1 thermal stressing in the temperature range 200 and 250 \ce{^0C}. Dimensions of heating tube are; length is 91 cm, tube outside diameter is 3.18 cm and inside diameter is 1.8 cm. Calculated volume for given heating tube dimensions is 2.314 ml. Flow tests samples thermally were stressed and collected at  flow rate 1 ml/min, pressure 600 psig (4.13 MPa) and temperatures at 200 \ce{^0C} and 250 \ce{^0C}. Static tests were conducted in the same apparatus at 600 psig (4.13 MPa) and 250 \ce{^0C} for 1 hour, 2 hours and 4 hours. Single tube flow reactors have been used for thermal stressing and thermal oxidative stability study of jet fuel as reported previously \cite{Corporan2011}.

Thermo Scientific Xactive mass spectrometer used for mass spectrometric analysis in this study is equipped with orbitrap detector and xcalibur software. Nitrogen as sheath gas was maintained at 15 psi and spray voltage was set at 4.5 kV. Auxiliary gas flow was not required for used sample flow rate (100 {\ce{\mu}l/min) and relatively clear background however it was set initially. Positive ion mode with electrospray ion source was applied for MS analysis since most species can be protonated by applying a positive potential gradient. Sample ionization in ESI source takes places by high applied voltage upto 5kV. Large molecules ionization without structural deformation or fragmentation is advantage of ESI over other methods of soft ionization \cite{Banerjee2012}. Noticeably, ESI-MS is a suitable method for mass spectroscopic analysis of relatively large molecular weight components in the mass range 300-1000 Da for hydrocarbon fuels and has been recently used for jet fuel analysis \cite{Commodo2011,Adams2013}. 
\\*

Methanol extraction of jet fuel samples was conducted in a separatory funnel and methanol to fuel ratio 4:1. Extracted fuel solvent volume layer was separated followed by further fuel dilution with solvent (fuel to methanol/IPA ratio, 1:100) for ESI-MS analysis. Intermittent washing steps were carried out after every sample run and ESI-MS spectrum recording. MeOH and water were used as solvent in ratios 50:50 and 60:40 in two methods with 0.1 \% v/v HCOOH for separation through isocratic elution. Second method with 60\% MeOH produced relatively better ionization and enhanced peak intensity for Jet A-1 samples. IPA and water were also used with second method for static test samples but ionization was better with methanol water solvent mixture.

Fourier transform infrared spectrometer instrument used for fuel samples analysis is Paragon 500 (spectral range 4000 cm-1 to 400 cm-1). Unstressed Jet A-1 samples, flow test sample at 1 ml/min and 300 \ce{^{0}C}, and 2 and 4 hour static tests samples at 250 \ce{^{0}C} spectra were recorded with FTIR. Jet A-1 NMR runs were conducted using Agilent DD 500 MHz instrument with carbon frequency 125.65 Hz. Distortionless Enhencement Polarization Transfer (DEPT) is a useful NMR technique which can differentiate between methyl (\ce{CH3}), methylene(\ce{CH2}) and methyne(CH) groups. Jet A-1 DEPT was recorder for $\theta$ = 135 which shows \ce{CH2} as inverted peaks and CH , \ce{CH3} as positive peaks.


\section{Results \& Discussion}

\subsection{Jet A-1 heteroatomic components}

Jet A-1 is a kerosene type aircraft gas turbine engine fuel. In this work, Jet A-1 heteroatomic, polar components and SMORS are analyzed by electrospray ion source mass spectrometry (ESI-MS), Fourier transfer infrared (FTIR) and \ce{^${13}$}C nuclear magnetic resonance spectroscopy (NMR). Higher molecular weight species are observed in Jet A-1 unstressed and thermally stressed samples spectra in the mass range 300-1000 Da. Jet A-1 mass spectra recorded heteroatomic hydrocarbon fuel components such as quinolines, indoles and SMORS as major species. Some background peaks are visible due to very high sensitivity of MS instrument, however with very low or negligible intensity compared to fuel components. In order to compare the effect of methanol extraction prior to MS analysis, mass spectrum of jet fuel sample not extracted with methanol is shown in figure \ref{fgr:Fig01} b and methanol extracted fuel sample mass spectrum in figure \ref{fgr:Fig01} a. Clearly, high relative abundance of polar heteroatomic fuel components in jet fuel mass spectrum in figure \ref{fgr:Fig01} a is because of methanol extraction prior to mass spectrometric analysis. Similarly, thermally stressed fuel samples extracted with methanol and IPA show high relative peak intensity in the mass range 200-1000 Da as shown in figures \ref{fgr:Fig02} a and b. Noticeably, SMORS such as m/z 589.4 are ionized better with MeOH extracted fuel samples and m/z 915.6 with IPA extracted thermally stressed fuel samples as shown in figures \ref{fgr:Fig02} . Jet A-1 higher molecular weight components  in mass range 350-1000 Da are m/z [M+H] 380.3, 437.1, 469.3, 483.3, 576.3, 589.9, 596.6 and 915.6. Major species with relatively high intensity and also as base peak(s) in mass spectra are 469.3, 589.9 and 915.6. 
\\*

\begin{figure}[h]
\centering
  \includegraphics[width=12cm,height=6cm]{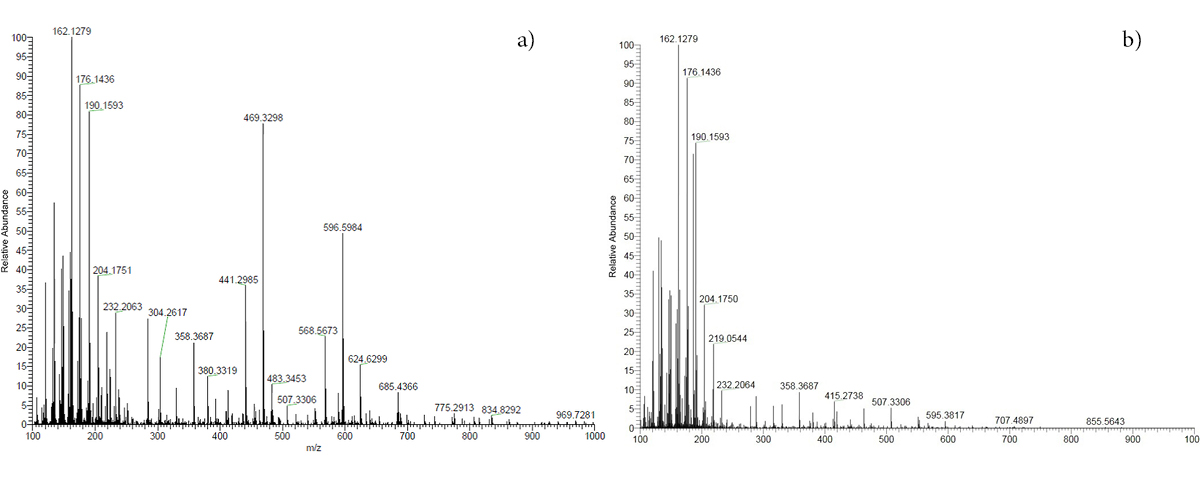}
  \caption{a) Electro spray ionization mass spectrum of methanol extracted unstressed jet fuel (Jet A-1), and b) Jet A-1 electro spray ionization mass spectrum without methanol extraction\label{fgr:Fig01}}
\end{figure}

\begin{figure}[h]
\centering
  \includegraphics[width=12cm,height=6cm]{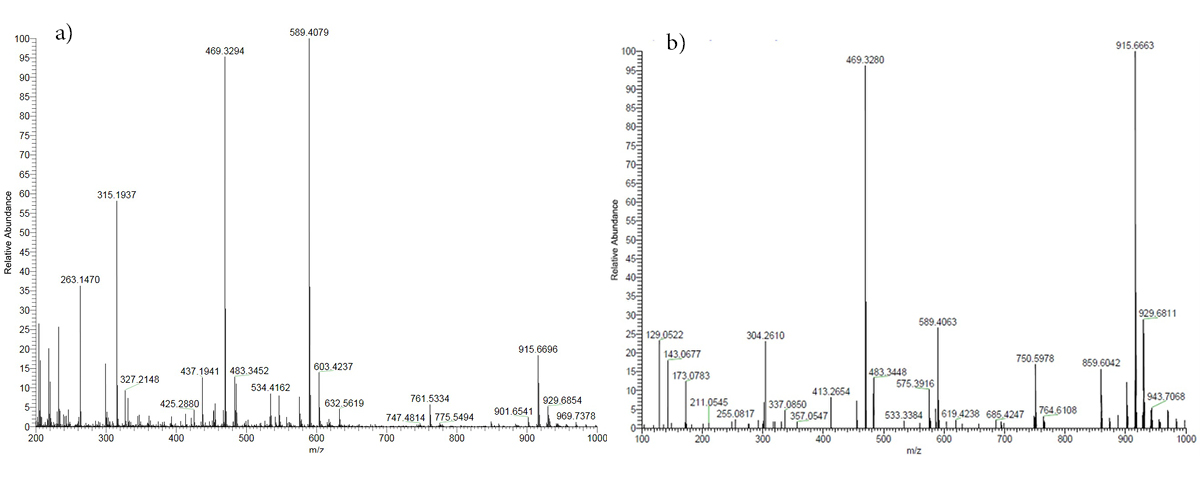}
  \caption{a) Electro spray ionization mass spectrum of methanol extracted thermally stressed four hours static test sample (Jet A-1), and b) IPA extracted thermally stressed two hour static test fuel mass spectrum with SMORS \label{fgr:Fig02}}
\end{figure}

Nitrogen containing aromatic compounds such as quinoline or isoquinoline correspond to the protonated ion 130.1. 2,6 dimethylquinoline is ionized with m/z 158.0 with moderate relative intensity. Software predicted NMR chemical shifts for N containing jet fuel components and Jet A-1 NMR are presented in the table (TableSI01) with supporting information accompanying this article. Few components such as m/z 173.1 and SMORS with m/z 589.4, 915.6 are thermally stressed fuel components and not observed in the unstressed fuel mass spectrum. FTIR spectra of Jet A-1 reveal absorption bands for alkanes, unsaturated and heteroatomic functional groups and few absorption bands are shown in figure \ref{fgr:Fig03}. Table \ref{tab:IRJetA1}  presents Jet A-1 FTIR absorption bands for both unstressed and thermally stressed samples. Preliminary NMR analysis conducted in this study revealed oxygen containing jet fuel components in the chemical shift range ($\delta$), 40-75 ppm with relatively low intensity as a result of their trace levels in the fuel. \ce{$^{13}$}C standard spectra of unstressed and thermally stressed fuel samples show Jet A-1 components and few alkoxy peak signals. Four hour static test samples shows increased peak intensity for $\delta$, 40-55 ppm compared with unstressed fuel spectrum in figures \ref{fgr:Fig04} a and b. Jet A-1 DEPT spectra were recorder for \ce{\theta} as 135 which show \ce{CH2} as inverted peaks and CH, \ce{CH3} as positive peaks. DEPT spectra for static and unstressed fuel samples are compared and show increased peak intensity for few peaks in 40-50 ppm in figures \ref{fgr:Fig04} c and d. Few negative peaks recorded in 40-75 ppm are 41.6, 54.3, 56.7, 63.0, 65.0, 65.6, 67.4, and 68.4. Clearly, increase in relative intensity in the alkoxy heteroatomic NMR chemical shifts can be associated with oxygen containing compound peaks in thermally stressed FTIR and SMORS observed in the fuel mass spectra.
\\*

\begin{figure}[h]
\centering
  \includegraphics[width=10cm, height=6cm]{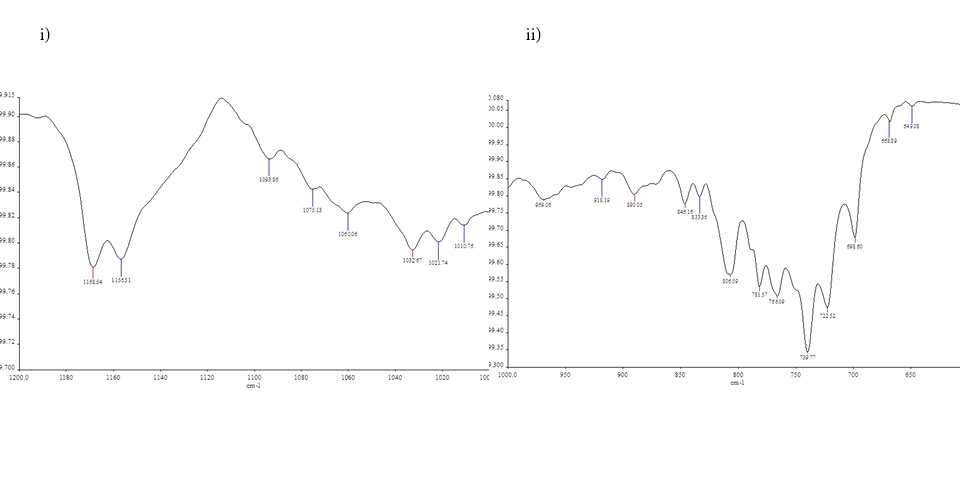}
  \caption{FIIR spectrum of Jet A-1, 1600 to 2000 \ce{cm^{-1}}\label{fgr:Fig03}}
\end{figure}

\begin{figure}[h]
\centering
  \includegraphics[width=12cm, height=9cm]{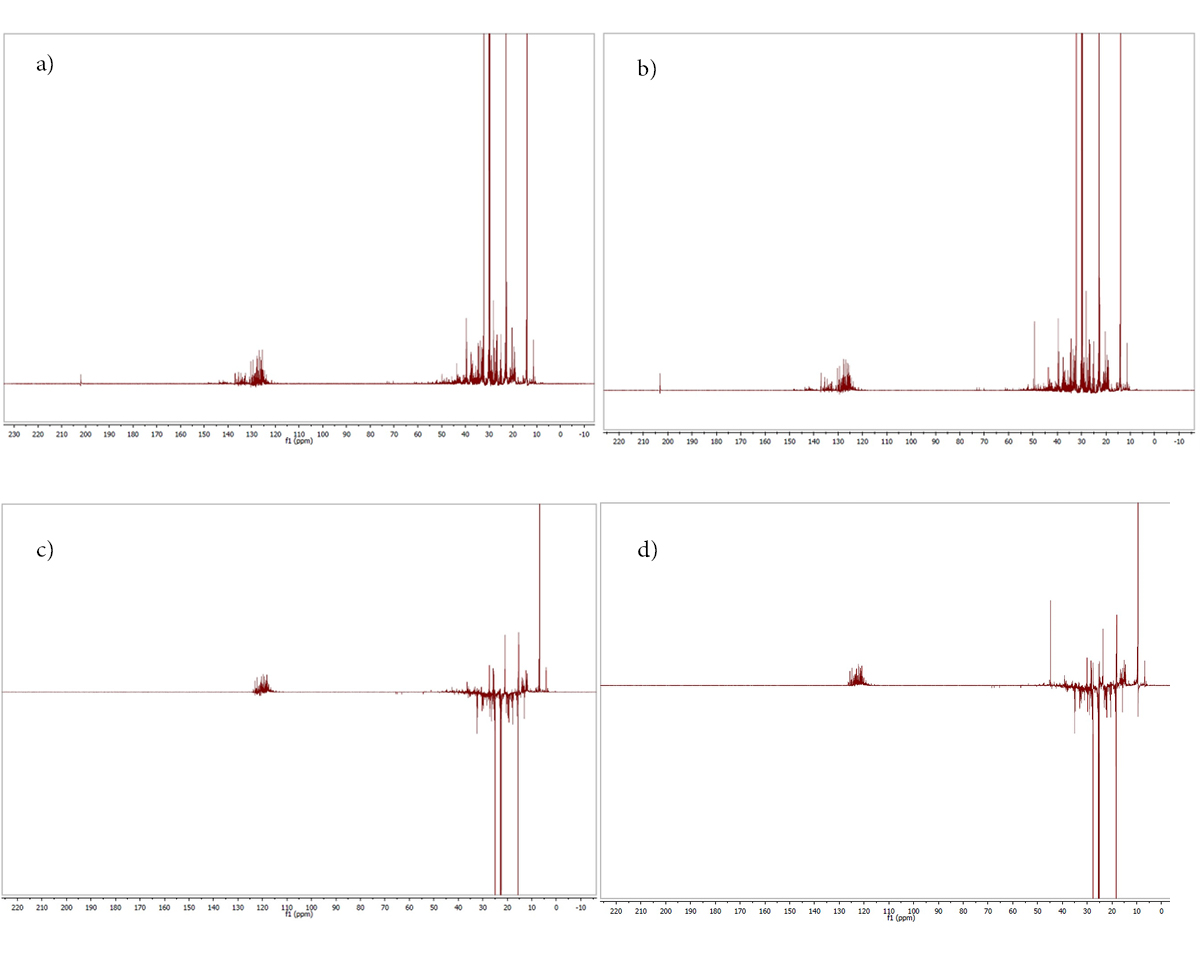}
  \caption{NMR spectra of unstressed Jet A-1 a) standard \ce{$^{13}$C} and c) DEPT 135; and thermally stressed four hour static test sample b) standard \ce{$^{13}$C} and d) DEPT 135\label{fgr:Fig04}}
\end{figure}

Nitrogen compounds such as indoles, quinolines and substituted pyridines promote deposition, and 2,5 DMP contributes more than other N containing species in the deposition \cite{Hazlett1991}.  In a recent study by Ryan et al \cite{Adams2013} series of heteroatomic compounds and possibly jet fuel species such as quinolines, indolines, tetrahydroquinolines, pyrroles and saturated indoles are reported as analyzed by positive ion mode ESI-MS. Jet fuel components in this study which are similar to above analysis\cite{Adams2013} and most likely as quinolines and amino naphthalene are m/z 144, 158, 172 and 186, indolines and tetrahydroquinolines with m/z 148, 162, 176 and 190, anilines and pyridines with protonated ions 150, 164, 178 and 192 and saturated indoles or pyrroles with m/z 180. Quinoline or isoquinoline and 2,6 dimethyl quinoline are N containing heteroatomic fuel components are identified in the Jet A-1 samples in this study. Oxygen containing functional groups are identified in the FTIR and NMR spectra in the fuel samples along with higher molecular weight polar components and SMORS. 
\\*

\begin{table}[htbp]
\caption{ Jet A-1 major components FTIR bands (\ce{cm^{-1}}) and oxygenated species vibrations with approximate descriptions\cite{Silver2004,Katrizky2000}}
\centering
\begin{tabular}{llc}\hline
Unstressed&Stressed&Approximate\\
fuel&fuel&description\\
 \ce{cm^{-1}}&\ce{cm^{-1}}&\\ \hline
&3734& OH, Free hydroxyl\\
2954&2954&\ce{CH3}, asym stretching\\
2924&2922&\ce{CH2}, asym stretching\\
2854&2854&\ce{CH2}, symm stretching\\
2732&2729&\ce{CH3}, symm stretching\\
&1932&\\
1457&1457&\ce{CH2}bending\\
1377&1377&\ce{CH2}, \ce{CH3}bending\\
1607&1607&Quadrant ring stretch\\

1739&1739&C=O $\alpha$,$\beta$ unsaturated esters\\
1170&1169&CH i.p.\\
&1156&C-O\\
&1093&C-O, ArOR\\
1077&1076&CHi.p.\\
1739&1739&C=O $\alpha$,$\beta$\\
&&unsaturated esters\\
&1682&C=O $\alpha$,$\beta$\\
&&unsaturated aldehydes, ketones\\
1077&1076,1075&CH i.p., COC, aryl alkyl\\
&1021&\\
&1010,1005\\
&971&Dienes, trienes trans-CH stretching\\
&918&OH bending\\
&890,889&\\
847&846&\\
833&833& CH o.o.p \\
810&806,808&$\beta$-naphthalene\\
782&781&\\
739&740,739&CH o.o.p\\
782&781&\\
770&766,766&CH o.o.p\\
739&740,739&\\
723&722,723&Dienes, trienes cis-CH stretching\\
699&698,699&OH phenyl o.o.p.\\
&668,648&\\

\end{tabular}
\label{tab:IRJetA1}
\end{table}

\subsection{Lichen substances as jet fuel component and potential additives}

Jet A-1 unstressed sample mass spectrum shows m/z 469.3 recorded in the positive ionization mode which corresponds to gyrophoric acid (GA) as shown in figure \ref{fgr:Fig01} a. Gyrophoric acid has molecular weight 468.4 and chemical formula \ce{C24H20O10}. Mass spectrum of GA by chemical ionization has been reported with a protonated compound peak at m/z 469 \cite{Addison1985} and was also identified in the mass spectrum of Lasallia Papulosa var rubiginosa \cite{Bohman1969}. Methanol and acetone are suitable extraction solvents for lichen substances. MS/MS spectrum in figure$\dag$ (FigureSI03)  provided as supporting information shows negligible fragmentation of m/z 469.3. MS/MS analysis was conducted with methanol extracted Jet A-1 flow test samples with high energy collision induced dissociation (HCD) ion mode on. FTIR analysis of Jet A-1 reveal few oxygen containing jet fuel components which are compared with GA infrared spectrum as reported by Huneck et al \cite{Huneck1996} and shown in table \ref{tab:LichenIR}. NMR spectra of Jet A-1 shows chemical shifts which correspond to GA NMR in table \ref{tab:Lichen13C}, reported by Huneck et al and Choudhary et al\cite{Huneck1996,Choudhary2011}. Carbon numbers assignment in the table \ref{tab:Lichen13C} for GA is in reference with reported GA NMR \cite{Huneck1996,Choudhary2011} and also provided in a figure as electronic supplementry information accompanying this article (FigureSI04). 
\\

\begin{figure}[h!]
\centering
\includegraphics[width=12cm,height=6.0 cm]{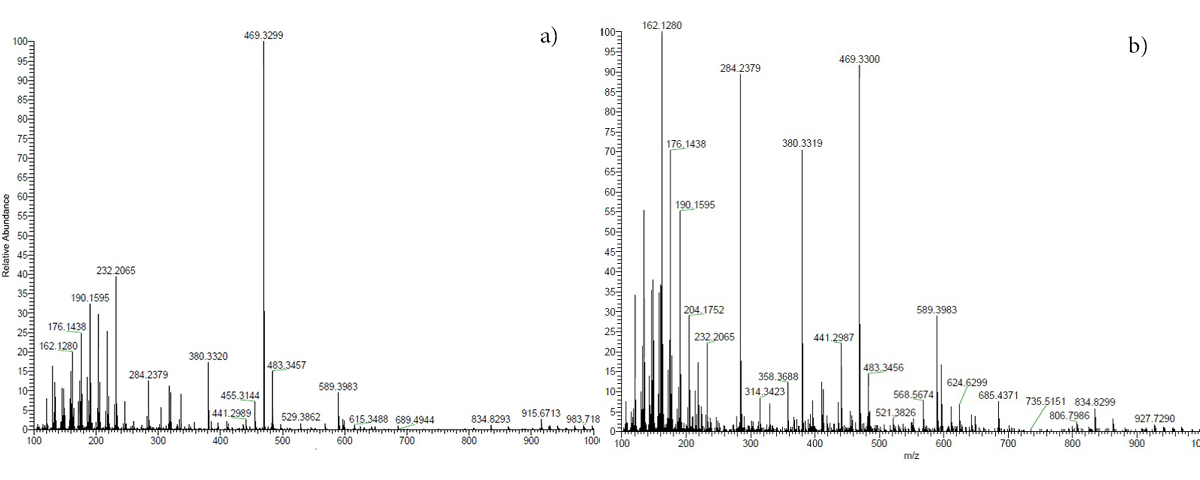}
\caption{Jet A-1 flow test methanol extracted mass spectrum with m/z 469.33 with high relative abundance}
\label{fgr:Fig05}
\end{figure}

\begin{figure}[h]
\centering
  \includegraphics[width=10cm, height=12cm]{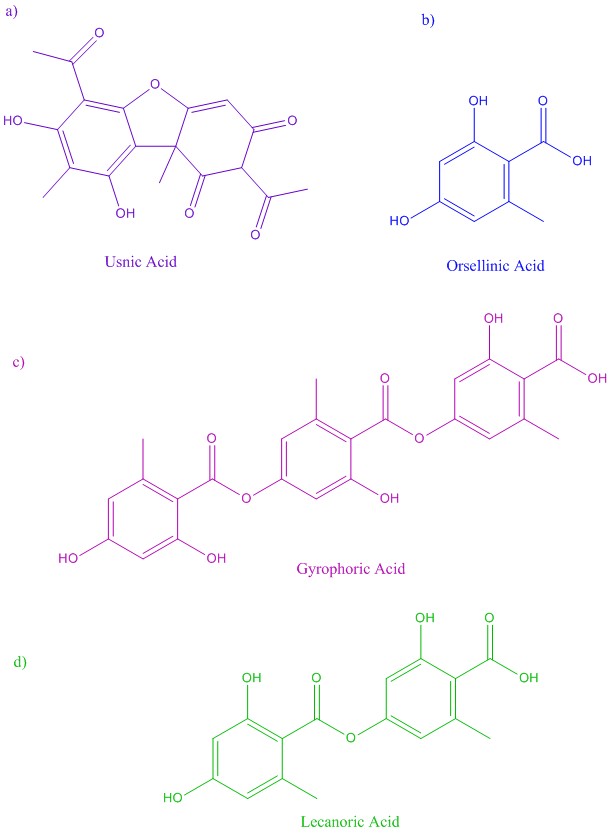}
  \caption{Gyrophoric acid a tridepside lichen substance as shown in c), a) Usnic acid, b) Orsellinic acid and d) Lecanoric acid \label{fgr:Fig06}}
\end{figure}

Thermal stressing of jet fuel was conducted by two methods; a) flow test and b) static test. Fuel was heated for short duration in the flow test while flowing in the heat exchanger tube at flow rate 1 ml/min. Deposition and SMORS formation in the Jet A-1 as a result of flow test is relatively less as shown in figure \ref{fgr:Fig05}. Static test was conducted for 4 hours with fuel stored inside the heat exchanger which generated higher molecular weight components such as m/z 589.4 and 915.6 with stable m/z 469.3 compound peak as shown in figures \ref{fgr:Fig02} a and b. GA or m/z 469.3 is ionized with high relative abundance and its peak intensity does not change significantly with heating at 250 \ce{^0C} during flow and static test. However, long duration thermal stressing results in SMORS formation such as m/z 589.4 and 915.6 with higher relative abundance than m/z 469.3.  Factors contributing to increased SMORS peak intensity in the mass spectra of static test are, first) longer duration of fuel thermal stressing, and second) decreased antioxidant activity of fuel additives at 250 \ce{^0C}. Although more investigation is required to confirm the antioxidant activity of GA during jet fuel thermal stressing, Jet A-1 MS analysis reveals m/z 469.3 as a relatively stable and possibly inhibiting oxidation in thermally stressed fuel except for 4 hour static tests suggests gyrophoric acid as Jet A-1 antioxidant. It is of course possible that antioxidant activity and deposition inhibition in the jet fuel samples can be due to other fuel additives. However, GA is an excellent antioxidant and antimicrobial agent as reported previously and possibly contributes in controlling oxidation while thermal stressing of Jet A-1.
\\

Gyrophoric acid (GA) is an antioxidant, antiproliferative agent and a tridepside produced by lichen found in old growth forests of North America such as lobaria linita, lobaria pulmonaria, lasallia papulosa and also produced by few other lichen species \cite{Kosanic2011}. Haemophaein is another dibenzofuran lichen substance with molecular weight same as GA. However, jet fuel infrared spectra reveal vibration bands corresponding to GA along with exhibiting antioxidant property as shown by mass spectra. Both tridepsides and dibenzofurans are produced by orsellinic acid cyclization as shown in figure SI05$\dag$. Tridepside GA isolated from umbilicaria sp. demonstrated the cytotoxic and antitumour activity  as reported by Burlando et al \cite{Burlando2009}. DPPH or di(phenyl)-(2,4,6-trinitrophenyl)iminoazanium is an antioxidant assay and GA exhibits a large antimicrobial and DPPH activity as an antioxidant\cite{Kosanic2011}. Natural products derived from plants and lichen also show excellent antioxidant and antiproliferative properties along with reduced harmful carcinogenic effect compared to standard antioxidants such as ascorbic acid, butylated hydroxyanisole (BHA) and butylated hydroxytoluene(BHT) \cite{Kosanic2011}. Also, lichen derivative GA which is found in lasallia pustulata has high phenolic content and exhibit free radical scavenging property \cite{Kosanic2011}. 
\\

\begin{table}[htbp]
\caption{ Lichen substance gyrophoric acid IR bands (\ce{cm^{-1}}) comparison with Jet A-1 spectra and approximate peak assignments \cite{Huneck1996,Silver2004}}
\centering
\begin{tabular}{l l c}\hline
Gyrophoric acid& Jet A-1&Approximate Vibration \\
\hline
3450, 3440&3734&OH\\
3150,3050,3067&&$\nu$(CH) aromatic\\ 
1690&1682&C=O, saturated ketone\\
1665,1640&&$\nu$(C=O) aromatic\\
&&conjugated carbonyls\\
1610&1607&Quardrant ring stretch\\
1574,1508&&\\
1385&1377&\ce{CH3} symm bending\\
1310&&\\
1240,1200&1217&COC, aryl,\\
&&alkyl ether asym stretching\\
1165&1169, 1168&CH i.p.\\
&1156&C-O\\
1140, 1132&&C-O\\
1050&1060,1032&\\
1000, 985&1010, 969&C-O\\
900&890&OH\\
870&&\\
840&846, 833&CH o.o.p.\\
740&739.7&CH o.o.p.\\
700&698&OH phenyl o.o.p.\\
\end{tabular}
\label{tab:LichenIR}
\end{table}

\begin{table}[htbp]
\caption{Gyrophoric acid 13C NMR chemical shifts $\delta$ (ppm) comparison with Jet A-1 spectra \cite{Huneck1996,Choudhary2011}}
\centering

\begin{tabular}{l l c}\hline

Gyrophoric acid& Jet A-1&Assignment\\ 
&&\\
C-4b: 155.0&152.0&C-OH\\
C-6,6b: 142,142&133.2,131.2&Quaternary C\\
C-6a:141.0&&\\
\\
C-5a,b: 116.0, 117.0&118.8&CH aromatic\\
C-1b,5: 114, 113.0&&\\
C-1a: 112.0&&\\
&&\\
C-3b: 109.0,C-1: 104.0&&\\
C-2: 102.0&&\\
C-8a: 22.3&22.3&\ce{CH3}\\
C-8b: 21.5&21.4&\ce{CH3}\\
C-8: 21.7&21.7&\ce{CH3}\\

\end{tabular}
\label{tab:Lichen13C}
\end{table}

 Lichen secondary metabolites such as depsides, depsidones and tridepsides are produced by acetate polymalonate pathway and orsellinic acid cyclization. Lichen species of division ascomycota and genus lobaria are lobaria linita and lobaria pulmonaria, and their figures (FigureSI01 and FigureSI02) are provided as supporting information with this article$\dag$}. Lobaria pulmonaria contains gyrophoric acid, tenuirin, constictic acid and norstictic acid. Lobaria linita is commonly known as lungwort or cabbage lungwort for its resemblance to lungs. Lobaraia linita and lobaria pulmonaria are lichens found in old growth forests of North America and are sources of orsellinate type of compounds such as tenuiorin, gyrophoric acid  (GA) and methyl gyrophorate \cite{Maass1975}.  GA is also identified in the extracts of lasallia papulosa which grows in the North American regions.  Lasallia papulosa belongs to the division ascomycota, class lecanoromycetes and family umbilicariaceae. GA is a tridepside  produced by lasallia papulosa \cite{Kosanic2011}, lobaria pulmonaria and lobaria linita \cite{Maass1975} and can be present in other lichen species. 
\\

Antioxidants prevent oxidation when present in low concentration or trace amounts compared with other oxidizable substrate.  Primary antioxidants are radical scavengers and secondary are peroxide decomposers. Commonly used fuel antioxidants and chemical additives are phenolic compounds such as 2,6 ditertbutyl phenols (DTB), butylated hydroxytoluene (BHT), aromatic amines, compounds containing sulphur and phosphorus. Few natural products antioxidants commercially available are gallic acid and ascorbic acid. Phenols are excellent antioxidants because of their acidic hydroxyl group and electron donating property. Phenols and phenoxy radical are resonance stabilized because of the presence of benzene ring. Phenoxide ion is more stable since it has only negative charge compared with phenol in which both positive and negative charges are present and result in less stability of phenol than phenoxide ion. Lichens are a potential source of phenolic compounds. Lichen products which contain oxygenated, aromatic and unsaturated rings are depsides, depsidones, diphenyl ethers and dibenzofurans. Depsides, tridepsides and tetradepsides consists of 2, 3 and 4 hydrobenzoic acid residues linked by esters. Depsidones arise by oxidative or orsellinic acid cyclization of depsides and reported as more efficient antioxidants \cite{Bucukoglu2013}. Orsellinic acid which is a monomolecular phenolic compound is the common base unit in lichen acids formed by orsellinic acid cyclization. Uniquely, lichen substances such as depsides, usnic acid, dibenzofurans are different from other plant derivatives and polyphenols by their characteristic acetate polymalonate derived aromatic phenols with 2 or 3 phenolic subunits linked through oxidative coupling and esterification \cite{Culberson1969}. Few chemical additives and lichen substances such as fatty acids and terpenes also have origin in other plants. Linoleic acid is one such compound and lichen substance which is found in many plants species and vegetable oils and is used as a jet fuel additive lubricity improver. Dimers and trimers of linoleic acid are also investigated in jet fuels \cite{Adams2013,Black1989}.
\\

Lichen acids  umbilicaric acid and gyrophoric acid exhibited antioxidants activities, however with relatively lower percentage compared to synthetic antioxidant butylated hydroxytoluene (BHT)\cite{Bucukoglu2013}. Umbilicaric acid and gyrophoric acid also exhibited excellent antibacterial activity against five gram negative bacteria \cite{Bucukoglu2013} including pseudomonas aeruginosa which is also a jet fuel contaminant. Another useful lichen acid is usnic acid which is an antioxidant, antimicrobial, and exhibit strong radical scavenging activity \cite{White2014, Rankovic2015}. Lichen acids lecanoric acid and salzinic acid are also antioxidant and antimicrobial agents. Additionally, synthetically produced antioxidants used as food and fuel additives such as butylated hydroxytoluene (BHT) can be toxic and carcinogenic. In contrast, lichen substances gyrophoric acid (GA) and usnic acid (UA) are reported antiproliferative, anticancer \cite{Kumar1999} compared with synthetic fuel additives. Extraction of lichen acids remains a key challenge, however few appropriate organic solvents commonly used for lichen acids purification are methanol, acetone and IPA. Recently, ionic liquids in combination with microwave assisted extraction has been reported as progressions in lichen metabolites extraction\cite{Bonny2011}. Optimization of lichen cultures and extraction of usnic acid with acetone and methanol is also reported by Behera et al \cite{Behera2009}.

 Oxygen containing species in jet fuel can reduce friction and wear due to their phenolic and carboxylic groups. For example, dicarboxylic acid esters are added to enhance lubrication properties of fuels to adapt the single fuel concept (SFC) introduced by National Atlantic Treaty Organization(NATO) \cite{Anastopoulos2013}. It is observed that more centered esteric linkage in a molecule provide better lubricity compared to similar ester containing compounds \cite{Wei1986,Anastopoulos2013}. Additionally, such species impact particulate emission with increased O/C ratio more than 0.2 \cite{Anastopoulos2013} and presence of esteric and other oxygen containing functional groups upto few ppm is desirable in jet fuels. Clearly, gyrophoric acid is a Jet A-1 component as identified in this work and potential middle distillate fuel additive for its multifaceted uses as antioxidant and antimicrobial properties along with esterically linked phenols structure suitable for kerosene type fuels.

\section{Conclusions}

In summary, heteroatomic Jet A-1 components and SMORS are identified by spectroscopic analysis in this study. SMORS are observed in the thermally stressed jet fuel samples. SMORS concentration depends on thermal stressing residence time and temperature, and increases with longer static tests compared to flow test. ESI-MS recorded heteroatomic fuel components and FTIR spectra revealed oxygenated compounds with very small bands corresponding to OH, C=O, COC, ArOR or aryl aklyl ethers. Jet A-1 NMR spectra also reveal chemical shifts indicating alkoxy species present in unstressed and thremally stressed fuel samples. 
\\

A lichen substance, gyrophoric acid (GA) is identified as a jet fuel component in this study. Excellent extractability of GA in methanol is advantageous for its easy extraction. GA shows negligible fragmentation in the mass spectrometric analysis as ionized as a pure compound  with antioxidant properties, hence is not a SMORS or fuel deposit component generated by thermal oxidative instabilities. ESI-MS, FTIR and NMR spectra of Jet A-1 show peaks corresponding with previously reported GA spectra. Lichens are abundant source of useful multifunctional polyphenolic additives. Advantages of lichen derived natural products are multifaceted with their anticancer, antioxidant and antimicrobial properties. Looking forward, advances in extraction processes can accelerate the use of  lichen treasured compounds as additives. 


\section*{Acknowledgements}
Author is thankful to Prof. \"Omer G\"ulder, University of Toronto Institute for Aerospace Studies, for his help and providing jet fuel samples for this study. 

\textbf{Supplementary information$\dag$} accompanies this article.
\\

\textbf{Competing financial interests}: Author declares no competing financial interests.

\end{document}